# *Chandra* Observations of Comets 8P/Tuttle and 17P/Holmes during Solar Minimum

D. J. Christian[1,2], D. Bodewits[3], C.M. Lisse[4], K. Dennerl[5], S. J. Wolk[6], H. Hsieh[7], T.H. Zurbuchen[8], and L. Zhao[8]

[1] Eureka Scientific, 2420 Delmer Ave, Suite 100, Oakland, CA, 94602, damian.christ@gmail.com

[2] Department of Physics and Astronomy, California State University Northridge, 18111 Nordoff Street, Northridge, CA 91330

[3] NASA Postdoctoral Fellow, NASA Goddard Space Flight Center, Solar System Exploration Division, Mailstop 690.3, Greenbelt, MD 20771, USA, Dennis.Bodewits@nasa.gov

[4] Planetary Exploration Group, Space Department, Johns Hopkins University Applied Physics Laboratory, 11100 Johns Hopkins Rd, Laurel, MD 20723, carey.lisse@jhuapl.edu

[5] Max-Planck-Institut für extraterrestrische Physik, Postfach 1603, 85740 Garching, Germany, kod@mpe.mpg.de

[6] *Chandra* X-Ray Center, Harvard-Smithsonian Center for Astrophysics, 60 Garden Street, Cambridge, MA, swolk@cfa.harvard.edu

[7] Queen's University Belfast, Department of Physics and Astronomy, Astronomy Research Centre, Belfast, UK, h.hseih@qub.c.uk

[8] Department of Atmospheric Oceanic and Space Sciences, University of Michigan, 2455 Hayward Street, Ann Arbor MI 48109-2143 thomasz@umich.edu; lzh@umich.edu

*Subject Heading*: comets:individual: (Comet 8P/Tuttle, Comet 17P/Holmes) - Sun:solar wind - techniques: spectroscopic - X-rays: General

# Abstract

We present results for *Chandra* X-ray Observatory observations of two comets made during the minimum of solar cycle 24. The two comets, 17P/Holmes (17P) and 8P/Tuttle (8P) were very different in their activity and geometry. 17P was observed for 30 ksec right after its major outburst, on 31 Oct 2007 (10:07 UT) and comet 8P/Tuttle was observed in 2008 January for 47 ksec. During the two *Chandra* observations, 17P was producing at least 100 times more water than 8P but was 2.2 times further away from the Sun. Also, 17P was at a relatively high solar latitude (+19.1 degrees) while 8P was observed at a lower solar latitude (3.4 degrees). The X-ray spectrum of 17P is unusually soft with little significant emission at energies above 500 eV. Depending on our choice of background, we derive a 300 to 1000 eV flux of $0.5 - 4.5 \times 10^{-13}$ ergs cm$^{-2}$ sec$^{-1}$, with over 90% of the emission in the 300 to 400 eV range. This corresponds to an X-ray luminosity between $0.4 - 3.3 \times 10^{15}$ erg sec$^{-1}$. However, we cannot distinguish between this significant excess emission and possible instrumental effects, such as incomplete charge transfer across the CCD. 17P is the first comet observed at high latitude during solar minimum. Its lack of X-rays in the 400 to 1000 eV range, in a simple picture, may be attributed to the polar solar wind, which is depleted in highly charged ions. 8P/Tuttle was much brighter, with an average count rate of 0.20 counts/s in the 300 to 1000 eV range. We derive an average X-ray flux in this range of $9.4 \times 10^{-13}$ ergs cm$^{-2}$ sec$^{-1}$ and an X-ray luminosity for the comet of $1.7 \times 10^{14}$ ergs sec$^{-1}$. The light curve showed a dramatic decrease in flux of over 60% between observations on January 1$^{st}$ and 4$^{th}$. When comparing outer regions of the coma to inner regions, its spectra showed a decrease in ratios of CVI/CV, OVIII/OVII, as predicted by recent solar wind charge exchange emission models. There are remarkable differences between the X-ray emission from these two comets, further demonstrating the qualities of cometary X-ray observations, and solar wind charge exchange emission in more general as a means of remote diagnostics of the interaction of astrophysical plasmas.

Proposed Running Title: ***Chandra*** **comet observations during solar minimum**

Please address all future correspondence, reviews, proofs, etc. to:

Dr. Damian J. Christian

damian.christian@csun.edu

# 1. Introduction

Comets in the inner solar system are generally bright X-ray emitters in the soft X-ray band (0.2 - 1.0 keV) with a total X-ray emitting power of 0.2-1.0 GW (Dennerl et al. 1997). The X-ray emission is driven by charge exchange between highly charged solar wind ions and the neutral gas in the coma (Cravens 1997; Krasnopolsky 1997). The charge exchange process is a quasi-resonant process, implying that the resulting emission is strongly dependent on properties of both the neutral gas (i.e. the comet) and the solar wind (Schwadron & Cravens 2000; Kharchenko & Dalgarno 2000, 2001; Beiersdorfer et al. 2003; Kharchenko et al 2003; Krasnopolsky 2004, 2006; Bodewits et al. 2004, 2006, 2007).

A recent survey of all comets thus far observed with *Chandra* (Bodewits et al 2007) demonstrated that the X-ray spectrum mainly reflects the state of the local solar wind during the observations. During solar minimum, the solar wind can be considered to be in a bimodal state, with a slow, highly variable wind within roughly 15 degrees of the Sun's equator, and a fast, tenuous and more steady polar wind at higher latitudes (McComas et al. 2003). These winds originate in different parts of the solar atmosphere, and the different temperatures at these source regions imply different charge balances. The slow equatorial wind stems from hotter streamer belt regions than the polar wind, that originates in polar coronal holes, and ions in the equatorial wind are on average in a higher ionization state than ions in the polar wind. During solar maximum, this bimodality disappears, leaving the solar wind in a highly chaotic state. As soon as charge exchange was acknowledged to be the driving process behind the cometary X-rays (Cravens 1997; Dennerl et al. 1997; Krasnopolsky 1997; Lisse et al. 2001), it was predicted that the bimodality of the solar wind should be reflected in cometary X-ray spectra (Schwadron & Cravens 2000; Kharchenko & Dalgarno 2001; Bodewits et al. 2004).

Two comparative reviews of all comets observed with *Chandra* in the period 2001-2006 (Bodewits et al. 2007; Krasnopolsky 2006) linked cometary spectra to solar wind states by comparing the low energy X-ray flux attributed to carbon and nitrogen ions with the higher energy X-ray flux attributed to oxygen ions. The derived ratio between the abundances of bare- and H-like oxygen ions (before charge exchange) was linked to solar wind freezing-in temperatures, showing that the *Chandra* comet sample covered a continuous range of solar wind states. Interestingly enough, the comets observed at high latitude were found to be interacting with hot, highly ionized solar winds. As these observations were all performed around 2001, these hot spectra could be linked to coronal mass ejections occurring around solar maximum. This led to the conclusion that *Chandra* had not yet observed a comet interacting with the cold, polar wind.

In this paper we present *Chandra* observations of two Jupiter family comets made during solar minimum: comet 17P/Holmes (hereafter 17P) and comet 8P/Tuttle (hereafter 8P). The observations of 17P were extraordinary due to both the state and the geometrical circumstances of the comet. Holmes is a Jupiter family comet that has underwent an unexpected major outburst that started around October 23.7 ± 0.2 (Schleicher et al. 2007; Gaillard et al. 2007; Sekanina 2008). While at a heliocentric distance of 2.44 AU and at 1.63 AU from Earth, the comet's visual brightness increased from $m_v$ = 17 to $m_v$ = 2.5. Whatever caused this outburst is yet unclear, but it released vast amounts of gas and dust (Montalto et al. 2008). According to SWAN observations by Combi et al. (2008), the water production reached a maximum of 1.4 x $10^{30}$ molecules $s^{-1}$ at Oct 27, and then decreased steadily in the weeks after. The comet remained a bright object on the sky for months after the outburst, mainly due to its exceptionally high dust-to-gas ratio. *Chandra* observed the comet a week after its outburst, on Oct 31. Observations done by Salyk et al. (2007) and Schleicher (2007) on the days before and after the observations indicate that around the time of our observations, the comet had a gas production rate between 3 – 5 x$10^{29}$ molecules $s^{-1}$. 17P thus had the largest gas production rate and heliocentric distance of any comet thus far observed with *Chandra*, and is the first comet that was observed with *Chandra* at high solar latitude (+19.1 degrees) during solar minimum.

Comet 8P/Tuttle is a comet from the Oort cloud reservoir now in a short-period Halley-type orbit, which has been suggested to be a contact binary (Harman et al. 2008; Lamy et al. 2008). It was observed as it passed within 0.25 AU of Earth in January 2008. 8P orbits the Sun with a period of 13.5 years with a high inclination (I = 55 deg), but during the Chandra observations on 1 – 4 January 2008 its heliographic latitude was 1 – 4 deg, i.e. very close to the Sun's equatorial plane. Water production rates were between $(2.1 – 2.4 \pm 0.1) \times 10^{28}$ molecules $s^{-1}$ on December 22 – 23 (Bonev et al. 2008) and $(1.4 \pm 0.3) \times 10^{28}$ on January 3, 2008 (Barber et al 2008).

17P and 8P represent the most recent comets observed by *Chandra*. These two comets are very different in their activity and geometry. During the two *Chandra* observations, 17P was producing at least 100 times more water than 8P but was 2.2 times further away from the Sun. 17P was the first comet observed by Chandra at a relatively high solar latitude (+19.1 degrees) while 8P was observed at a lower solar latitude (+3.4 degrees). 8P shows X-ray morphology and an emission spectrum that are more typical for comets, and is also a bright comet with which to test spectral changes as a function of distance from the nucleus predicted by current solar wind charge exchange (SWCX) models (Bodewits et al. 2007). This paper focuses on the observational properties of each comet and comparison to the simple SWCX model. A future paper will investigate further details of the CXE model under a Chandra archival proposal (program CXO-09100455). The outline of this paper is the following; Section 2 presents the observations and analysis. Imaging and temporal results are presented in Sections 3 and 4, respectively. In section 5 we discuss the total spectra, and for 8P we analyze spectra as a function of position from the nucleus, and spectra as a function of time. Lastly, in section 6 we discuss the results and compare them to solar wind conditions, previous comet observations and current SWCX models.

# 2. Observations & Analysis

**2.1 Observations**

Comet 17P was observed on 31.42 October 2007 for 30 ksec as part of a DDT with PI K. Dennerl. Each pointing obtained ~10 ksec on the comet for a total integration time of 30 ksec. Comet 8P was observed as part of a cycle 9 GO program (9100452; PI Christian) between 2008 January 1 07:31 UT and 2008 January 4 04:55 UT. Nearly 47 ksec of data were obtained in 3 observation blocks during each of 5 separate pointings on January 1, 3, and 4, 2008. Details of the observations and the observing geometry are summarized in Table 1.

The *Chandra* observations were done in the same manner as for our previous comet observations (Lisse et al. 2001, 2005; Bodewits et al. 2007; Wolk et al. 2009) and the reader is referred to these papers for details of the observational scheme. Briefly, the ACIS-S CCD was used, providing high resolution X-ray imaging with a plate scale of 0.5″ pixel$^{-1}$, an instantaneous field of view 8.3´ x 8.3´, and moderate resolution spectra ($\Delta E \sim 110$ eV FWHM, $\sigma_{Gaussian} \sim 50$ eV) in the 300 – 2500 eV energy range. The observations were conducted with the comet's nucleus near the aim-point in the S3 chip. No active guiding on the comet was attempted, and *Chandra* was able to follow the comet's motion using multiple pointings. In this method, the comet is centered in the S3 chip and allowed to drift across the chip and the *Chandra* pointing is updated to re-center the comet as before it moves off the chip edge.

**2.2 Analysis**

The *Chandra* data reduction and analysis were done in the same manner as for our previous comet observations (Lisse et al. 2001, 2005; Bodewits et al. 2007; Wolk et al. 2009) and the reader is referred to these papers for additional details. We provide only a summary of the analysis here. For each comet, combined images from each set of observations were remapped into a coordinate system moving with the comet using the 'sso_freeze' algorithm, part of the *Chandra* Interactive Analysis of

Observations (CIAO) software (Fruscione et al. 2006). Images and spectra were extracted with the CIAO tools and analyzed with a combination of IDL and FTOOL packages. The X-ray spectral analysis software package XSPEC (Arnaud 1996) was used for the spectral fitting. For each comet, we extracted spectra and associated calibration products (response matrices and effective areas) using the current *Chandra* Interactive Analysis of Observations tools (CIAO v4).

**2.2.1 17P**

The resulting effective field of view of the observations in the comet-centric coordinate system was about 9.3´ x 8.6´, and a total of approximately 7100 photons were detected in the 250 to 1000 eV range (of which 5300 were between 250 to 500 eV) on the entire S3 chip. Thus, the average total count rate in the 250 to 1000 eV energy range over the 30 ksec of observations was 0.24 counts/s. We extracted spectra from several apertures along the top of the S3 chip and from the entire S3 chip in comet-centered coordinates. The entire S3 chip had ~60% more counts than a rectangular aperture of 1100x550 pixels along the top of S3 (this region is near the CCD's readout nodes). Because of this fact, along with the large extent of the comet in the optical (much larger than Chandra's field of view), we concentrated our analysis on the spectrum extracted from the entire S3 chip. A background spectrum of the same dimensions was extracted from the bottom of the S3 chip.

The expected cometary position was verified with several other catalogued X-ray sources: 1RXS J034619.3+504139 was detected in the I2 chip, an un-catalogued source was found on the S3/S4 boundary at 03:47:17 +50:29:51, and star TYC 3338-277-1 (Voges et al 1999) was found in the S1 chip.

Ground and space borne optical observations of 17P such as obtained with SuperWASP (Pollacco et al. 2006; Hseih et al. 2007) show that around Oct 31, the coma had reached a diameter larger than 45', considerably larger than the ~8' x 8' ACIS-S3 FOV (see Figure 1). The comet therefore extended beyond the S3 chip, but because chips I2 ad I3 and S2 and S4 are front illuminated and have effectively no sensitivity below ~500 eV, 17P was not detected in these. Because the extent of the comet was larger than the ACIS field of view, we investigated other background spectra in addition to the spectra extracted from the lower region of the S3 chip. We searched the *Chandra* archive for ACIS-S observations at similar Galactic latitudes that would have a similar cosmic low energy background, see Table 2. The choice of background is not trivial and has significant consequences for our analysis. It will therefore be treated in more detail in section 5.1.

### 2.2.2 8P

The effective field of view of the observations in the comet-centric coordinate system was 17.6´ x 9.2´, in which a total of approximately 9330 cometary photons detected in the 300 to 1000 eV range over the entire S3 chip. Thus, the average net count rate in the 47 ksec of observations was 0.25 counts/s. The re-mapped 8P comet image in the 300 to 1000 eV range is shown in Figure 2. With 8P's discernable image morphology, we extracted spectra from several apertures centered on the comet and from the entire S3 chip. Background spectra were also extracted from several regions away from the comet and a spectrum extracted from the top and bottom of the S3 chip was chosen as having little cometary emission. These regions are: A.) upper and lower background apertures at the top and bottom of the S3 chip, B.) a circular aperture of 100 pixels centered on the brightest central emission and labeled *nucleus*, C.) 3 concentric semi-circles, noted as *pandas* 0,1, and 2 centered on the brightest point of the nucleus and arranged toward the Sun-ward side, and D.) a semi-circular region on the anti-Sun side of the nucleus labeled *left-panda*. We also extracted full-chip spectra as a function of time, as the comet's X-ray emission decreased over 60% during the observation (see Section 4).

# 3. Imaging

**3.1.1 17P**

The reconstructed comet image is shown in Figure 1 and is compared to the wide field optical image from the SuperWASP project (Pollacco et al. 2006) in the V-band. Examining the ACIS S3 chip in the usual 300 – 1000 eV range did not show an obvious comet or a strong concentration of photons. However, the X-ray image shows the highest concentration of X-ray photons in the top part and upper left region of the S3 chip. This emission is more clearly visible when examining the 300 to 400 eV energy range, and this image is shown in Figure 3a. The central emission does appear to have 2 peaks and an extension to the right rather than any recognizable morphology. These two regions have an excess of ~40 counts over background in the 300-400 eV range, or a detection of ~6$\sigma$.

We then examined several background images (Table 2, section 5.1) in the 300 – 400 eV range to test if the excess soft emission could be an instrumental effect. For the ID 8473 S3 image we also find an excess emission in the 300 – 400 eV range. The 17P and ID 8473 images are compared in detector coordinates in Figure 3b-d. This soft excess cannot be easily distinguished from excess caused as resulting from charge transfer inefficiency of photons in the 300 – 400 eV range. The charge from these photons does not propagate efficiently across the chip and are preferentially seen closer to the readout end of the chip. There may still be a small excess of soft energy photons from 17P visible in the image of 17P after background subtraction with the ID8473 image. This image is also shown in Figure 3d, and an excess, 40 counts (above background) in a circle with 100 pixel radius, can be seen in the lower right hand side.

**3.1.2 8P**

The reconstructed comet image of 8P is shown in Figure 2 and shows the typical crescent-shaped emission, similar to other comets, such as Hyakutake (Lisse et al. 1996), Linear S4 (Lisse et al 2001), and C/2001 WM1 (Wegmann & Dennerl 2005). The 2-dimensional image morphology is very similar to model predictions (Wegmann et al. 2004; Wegmann & Dennerl 2005, Bodewits et al. 2006). 8P, with its low gas production rate in the order of $2 \times 10^{28}$ molecules sec$^{-1}$ (Bonev et al. 2008; Barber et al 2008) has a morphology very similar to the low gas production rate model of 2P/Encke viewed at a high phase angle (e.g., see Wegmann et al. 2004, Figure 4d).

## 4. Temporal Behavior

### 4.1.1 17P

X-ray light curves due to charge exchange emission are a combination of the comet's gas production and the solar wind heavy ion flux. The X-ray light curve for 17P is constant for the entire 30 ksec observation with a net average count rate of $0.05 \pm 0.01$ cps for our nominal background. The low count rates do not allow for a more detailed analysis of the development of the spectrum over time. Because the comet had a very large gas production rate during the observations ($3 - 5 \times 10^{29}$ molecules s$^{-1}$), it must have been collisionally thick to charge exchange. The X-ray emission than mainly comes from the outer parts of the coma, which is dominated by fragment species (rather than gases released directly from the nucleus). Short term variations will consequently be dominated by variations in the solar wind. Based on the observed constant rate average count rate, it is to be expected that 17P interacted with a quiet solar wind with no large variations in its (heavy ion) composition or flux.

### 4.1.2 8P

The 8P light curve shows a large decrease in counts between the first of the observations on Jan 1 (0.4 counts/s) to Jan 4 (0.10 counts/s). We show the *Chandra* 300 - 1000 eV light curve in Figure 4. A small flare was observed on January 1$^{st}$, but the overall trend is a decrease in count rate. 8P had a much smaller gas production rate than 17P and most of the coma will have therefore been collisionally thin to charge exchange (Bodewits et al 2007), allowing the solar wind ions to interact with parent species close to the nucleus. X-ray brightness variations can thus indicate both variations in the gas production, and in the solar wind. Short term spectral variations however might indicate solar wind rather than cometary variations. Searches for periodic signals in the CXO light curve indicative of the comet's rotation period (between 5.7 and 7.4 hours; Woodney & Schleicher 2008; Harmon et al. 2008) were inconclusive as there is relatively poor phase coverage and count rates are dominated by the higher count rate data on January 1st. We investigate 8P's emission spectra in the following section.

## 5. Spectroscopy

The majority of prominent X-ray emission lines can be attributed to the solar wind highly charged ions of $C^{5+}$, $C^{6+}$, $N^{6+}$, $N^{7+}$ and $O^{7+}$, $O^{8+}$ and to avoid confusion, we denote the emission line formed with a roman numeral, e.g., if an $O^{7+}$ ion captures and electron it results in an O VII emission line near 561 eV. Spectral models for cometary X-ray emission have improved significantly since earlier simple continuum models, such as thermal Bremsstrahlung fit to broadband measurements (e.g, Lisse et al. 1999). For the cometary X-ray spectra presented here we have used the SWCX model of Bodewits et al. (2007). In this SWCX model, each group of ions in a species were fixed according to their velocity dependent emission cross sections to the ion with the highest cross section in that group. Thus, the free parameters were the relative strengths of H- and He-like Carbon (299 and 367 eV), Nitrogen (N V at 419 and N VI at 500 eV) and Oxygen (O VII at 561 and O VIII at 653 eV). Additional Ne lines, the Ne IX forbidden, inter-combination and resonance lines around 907 eV and the Ne X Ly-alpha at 1024 eV were also included in our SWCX model, giving a total of 8 free parameters.

**5.1 17P**

The extracted S3 full-chip spectrum was very soft, with almost no emission from the OVII and OVIII features in the 500 to 700 eV range that are typical for comet SWCX emission. We used several different background spectra to examine 17P's X-ray emission. Firstly, we used a region from the lower part of the S3 chip. Secondly, we selected a background spectrum from a 120 ksec observation of SN1986J (ID4613) which was observed in December 2003 but was the closest match to 17P's Galactic position. Thirdly, we used a 30 ksec observation of SN2006gy (ID 8473), which also had a similar Galactic position to 17P and did not show any resolved point sources in the S3 image. And lastly, we used an ACIS 450 ksec blank sky observation as a background spectrum. Other archival background spectra with shorter exposure times, and hence lower signal-to-noise produced similar fluxes for the background subtracted 17P spectrum. The 450 ksec ACIS blank sky spectrum had a very different shape and greatly over-subtracted the 500-700 eV region. We attribute this to the fact that the ACIS blank sky observations are a compilation of several different pointings and observing epochs. Thus it is a mixture of different X-ray background with much larger OVII/VIII components and we did not perform further analysis with this background.

The total full chip spectrum (no background removed) is shown in Figure 5a. Using the background spectrum from the bottom of the S3 chip leaves almost no flux below 300 eV and a soft spectrum with almost all of the emission in the 300 to 400 eV range (see Figure 5b). Because the background region is closer to the CCD read-out the 250 - 300 eV photons are preferentially higher in this region than the source region from the far side of the detector, thus the background subtracted spectrum is negative below 300 eV.

A comparison between the total signal from the S3 full chip and the background spectrum extracted from observation ID4613 (after removal of point sources) is shown in Figure 5c. This background spectrum had a 200 - 400 eV flux similar to that attributed to 17P and a similar lack of emission at energies greater than 500 eV.

Lastly, we also used a background extracted from the entire S3 chip of the SN2006gy (ID8473) observation in detector coordinates. This spectrum has a very similar 200-400 eV flux to 17P, but has a much larger excess at energies greater than 500 eV, probably from the soft X-ray background (see

Figure 5d). Thus the background subtracted 17P spectrum is negative for fluxes above approximately 450 eV and only shows a small positive excess in the 250-400 eV range.

We then fit the 17P spectrum with each of the 3 backgrounds using the solar wind charge exchange (SWCX) model of Bodewits et al. (2007). The spectral parameters are given in Table 4 and the spectrum with model components is shown in Figure 6. We find a flux of $4.3 \pm 0.7 \times 10^{-13}$ ergs cm$^{-2}$ sec$^{-1}$ in the 300 to 1000 eV range using the 17P S3 lower background, a flux of $4.5 \pm 0.7 \times 10^{-13}$ ergs cm$^{-2}$ sec$^{-1}$ for the ID 4613 background, and a flux of flux of $0.5 \pm 0.3 \times 10^{-13}$ ergs cm$^{-2}$ sec$^{-1}$ for the ID 8473 background. We note over 90% of the observed flux is in the 300 to 500 eV range, but give the 300 to 1000 eV flux for comparison to pervious studies (e.g. Bodewits *et al.* 2007).

We also tested the effect of increasing the background on 17P's spectrum. Increasing the background about 50% removed all comet flux above about 400 eV and the 300 to 400 eV spectrum can be fit with only 2 emission lines, CV at 299 eV and CVI at 367 eV giving a reduced $\chi^2$ of 0.7. The derived flux for this increased background was $1.6 \pm 0.3 \times 10^{-13}$ ergs cm$^{-2}$ sec$^{-1}$, and thus it has 2/3$^{rd}$ less flux than the nominal background subtraction. Clearly we are over-subtracting the background with this scaling, but it does give us an additional lower limit to the flux.

**5.2 8P**

The 8P spectrum shows the typical strong O VII and moderate C V, C VI, N VI, N VII, Ne IX, and Ne X emission. The fitted SWCX model had a reduced $\chi^2$ = 1.2 and is a reasonably good fit for 47 degrees of freedom. These spectral parameters are given in Table 5 and the 8P full-chip spectrum is shown in Figure 7. We find a flux of $9.4 \pm 0.8 \times 10^{-13}$ ergs cm$^{-2}$ sec$^{-1}$ in the 300 to 1000 eV range. This gives an X-ray luminosity of $1.7 \pm 0.1 \times 10^{14}$ ergs sec$^{-1}$ in this energy range.

Spectra extracted as a function of distance from the comet's nucleus and as a function of time during the observation showed remarkable changes in the carbon and oxygen emission. Spectra at distances closer to the comet's nucleus showed a decrease in the OVIII flux consistent with SWCX models. The spectral fits to these regions are summarized in Table 5 and the spectra are shown in Figure 8. Spectra extracted as a function of time showed a dramatic change from the brightest interval (A) on January 1$^{st}$ to the faintest interval on January 4$^{th}$. On January 1$^{st}$, the oxygen (OVII + OVIII) emission is approximately 2 times stronger than the combined low energy 300 – 500 eV emission during the highest count rates and decreases to about 50% greater on January 2$^{nd}$, to being about equal on January 4$^{th}$. The January 4$^{th}$ spectrum has very weak overall emission and is approximately 60% fainter than the January first emission. The small flare spectrum on January 1$^{st}$ has a very similar shape to the overall spectrum on that day, but about 15% more flux. The spectral fits to these regions also summarized in Table 6 and the spectra are shown in Figure 9. We compare these trends in both distance from the nucleus and time to the current SWCX models in Section 6.

## 6. Discussion

### 6.1 Did Chandra Detect 17P?

In the preceding sections, we described how we searched for signatures of X-rays from comet 17P by examining both the images and spectra. Using a suitable background from the lower part of the S3 chip or from a similar Galactic position (ID4613) we derive flux of $4.5 \pm 0.7 \times 10^{-13}$ ergs cm$^{-2}$ sec$^{-1}$. Much of this emission may be the result of charge transfer inefficiency, which artificially enhances emission in the 200 to 400 eV range from the side of the detector furthest from the read-out amplifier. Subtracting a spectrum in detector coordinates from ID 8473 gives a flux of only $0.5 \pm 0.3 \times 10^{-13}$ ergs cm$^{-2}$ sec$^{-1}$ and this may be the true emission from the comet. These fluxes give a range of luminosities for 17P of $0.36 – 3.3 \times 10^{15}$ ergs sec$^{-1}$. Neither method yielded conclusive evidence of cometary X-rays from 17P during our 30 ksec *Chandra* observation

One possible explanation for the lack of image morphology could be the large extent of the coma due to the comet's high gas production rate and its large heliocentric distance. If the coma were collisionally thick to charge exchange outside the the S3 detector's 8.3′ x 8.3′ field of view, this could explain the extremely soft X-ray spectrum as charge exchange reactions decrease the average charge of solar wind ions along their trajectory through the gas around the comet, and will produce softer X-ray emission in subsequent reactions. At the distance of 17P, 8.3' corresponds to approximately $5.8 \times 10^5$ km. According to the comet-solar wind models that we discussed in our survey paper (Bodewits et al 2007), it is most unlikely that there is a notable collisional opacity effect on the ionic flux ratios at these distances from the nucleus, even for large gas production rates (see e.g. Fig. 7 in Bodewits et al. 2007).

We note that *Chandra* ACIS-S is not very well suited for study of these soft energies. If we believe we did detect 17P near the outer edge of the S3 chip we then have to explain how the X-rays can originate ~3.5′ or ~175,000 km from the nominal nucleus. The brightest portion of the emission in the re-mapped coordinates is in the comet-Sun direction . Optical studies of 17P suggest a spherically symmetric gas distribution, in which case the X-ray emission would be concentrated on the Sun-ward side of the coma along the comet-Sun axis. Uncertainties from out-gassing or other dynamical effects may therefore be needed to explain why the comet X-rays are a large distance from the nominal nucleus, but there is a significant detection above background in the 300-400 eV energy range.

Due to the high latitude and large geocentric distance of 17P, it is unreasonable to assume solar wind conditions measured by any of the spacecraft located at L1. The Ulysses spacecraft was positioned at a higher latitude and larger heliocentric distance. Around our observations, it measured fast coronal hole wind, but the variations do not allow for a simple extrapolation to the position of the comet.

## 6.2 Spectral changes of 8P

Spectra extracted as a function of distance from the nucleus do show an interesting trend that the ratios of CVI/CV, OVIII/OVII decrease when comparing outer regions of the coma to inner regions.. These ratios are shown in Figure 10a. These observed changes are expected in the SWCX model as solar wind ions lose their charge capturing electrons from the cometary neutrals (Bodewits 2007). The CVI/CV and NVII/NVI ratios also decreases about a factor of 2 from the outer to inner regions, although no strong difference is observed between the 2 inner regions (panda 0 and panda1).

Spectra extracted as a function of time, as 8P showed a 60% decrease in count rate, showed that the spectrum flattened and the usual ratio of oxygen emission being 50% larger than the C+N changes to be nearly equal for the lowest count rates. Similarly, the ratios of OVIII/OVII became more equal from there values of 0.3 at the start of the observations. These ratios for the spectra extracted as a function of time are shown in Figure 10b. Thus, C V and O VII emission dominates 8P's spectrum during the brightest and most flare-like intervals.

To compare our observations with solar wind conditions measured around L1, we applied a simple solar wind extrapolation scheme, outlined and discussed in our survey paper (Bodewits et al. 2007). In brief, one assumes constant outflow and tangential velocities and calculates the time lag or lead between spacecraft and comet to get an idea about the solar wind conditions resulting in the emitted X-rays. This extrapolation is a strong simplification of solar wind progression, and is only of use in relatively stable solar wind conditions, and when the comet is close to Earth. Further details of this procedure are also outlined in Neugebauer, M. et al 2000. We obtained estimates of the solar wind conditions at comet 8P during the Chandra observations from SOHO and ACE spacecrafts, shown in Figure 11. From these solar wind data, we observe that 8P interacted with a quiet, slow solar wind moving at approximately 300 to 400 km/s. The solar wind on January 1st reached nearly 400 km/s as observed in the ACE ion and SOHO proton velocities data. The wind velocity decreased to below 300 km/s on the 4$^{th}$, and the decrease in the Chandra count rate reflects this. We compare the 8P X-ray light curve and the solar wind data in Figure 11. The $O^{7+}/O^{6+}$ ion data shows minimal variation and is always below 0.2 during the observation, indicating a steady solar wind with a constant freeze-in temperature, but does show a slight increase near the observed flare on January 1$^{st}$.

## 6.3 17P and 8P and other comets

We derived a range of luminosities for 17P of $0.36 – 3.3 \times 10^{15}$ ergs sec$^{-1}$, and found 8P's luminosity to be $1.7 \times 10^{14}$ ergs sec$^{-1}$. Although our nominal luminosity of 17P ($3.3 \times 10^{15}$ ergs sec$^{-1}$) is on the lower end for X-ray luminosities of comets (see Figure 12 in Wolk et al. 2009), it is concentrated entirely in the 300-400 eV range. 8P was a relatively faint X-ray comet and only 2P/Encke and 73P/Schwassmann-Wachmann 3's fragment B were less luminous in X-rays than 8P. These three comets all had comparable gas production rates (~$10^{28}$ molecules/s) and were observed during moderate solar wind conditions. If not for the solar outburst on January 1$^{st}$, 8P would have been even 3 times fainter.

Both comets encountered unique solar wind conditions. Solar wind condition for 8P and 17P and a comparison to several of the comets from the previous Chandra survey (Bodewits et al. 2007) are given in Table 7. We include Comet C/1999 S4 (Linear S4, LS4), which encountered a ~600 km/s wind and has been a well studied comet to use for comparison. We also included Comet C/1999 T1 (Mcaught-Hartley, hereafter McH), which also encountered a solar wind designated as having a "hot" composition and with a bulk velocity of 353 km/s (Bodewits et al. 2007). Previously, we had derived solar wind abundances relative to $O^{7+}$, but since 17P had no $O^{7+}$ emission, we normalized its carbon and nitrogen emission to that of comet LINEAR S4 (hereafter LS4) and show these in Table 8. 17P encountered both a slow 400 km/s wind and a 700 km/s wind. Its very soft X-ray spectrum, with little emission above 400 eV is consistent with predictions made by Bodewits et al. 2007 for a very soft X-ray spectrum from a fast, cool high latitude wind. 17P has a much lower $C^{5+}$ abundance than LS4, but its large $C^{6+}$ ratio (38) is similar to comets that encountered solar winds with corotating interaction regions (CIRs) or flare events, consistent with the Ulyssess solar wind observations. The $N^{6+}$ ratio is much lower as expected for the soft spectrum of 17P and $Ne^{9+}$ and $Ne^{10+}$ values are limited by poor statistics.

8P encountered a relatively slow solar wind of ~360 km/s and its solar wind conditions are also summarized in Table 7. Its $O^{8+}$ abundances for both the total and flare spectra are comparable to those of LS4 and McH, although slightly lower. 8P's $C^{6+}$ abundance is slightly higher than the LS4 values and nearly 50% larger than he McH value. However, 8P has much larger $C^{5+}$, with a value of $\approx 70$, which is more comparable to comets which encountered a solar wind that included a flare or coronal mass ejection (CME). 8P's $N^{6+}$ and $N^{7+}$ abundances are also much larger than LS4 or McH's values are again more consistent with flare or CIR solar wind conditions. . 8P's $Ne^{10+}$ abundance is also larger, but limited by poor statistics.

# 7. Conclusions

We have presented results for the two most recent *Chandra* ACIS-S observations of comets made during solar minimum. Comet 17P/Holmes X-ray spectrum is unusually soft with little significant emission at energies above 500 eV. After careful background subtraction we derive a 300 to 1000 eV flux of  $4.5 \times 10^{-13}$ ergs cm$^{-2}$ sec$^{-1}$, with over 90% of the emission in the 300 to 400 eV range. The comet's X-rays are 3.5′ from the nominal nucleus, and uncertainties in the comet's position from out-gassing or other dynamical effects may be needed to explain this. However, we cannot distinguish between this significant excess emission and possible instrumental effects, such as incomplete charge transfer across the CCD. The lack of X-rays in the 400 to 1000 eV range may be attributed to the high latitude of 17P and its interaction with the polar solar wind which is depleted in highly charged ions in accordance with model predictions for SWCX at high latitudes (Bodewits et al. 2007). 8P/Tuttle was much brighter than 17P, with an average count rate of 0.20 counts/s in the 300 to 1000 eV range. We derive an average X-ray flux in this range of 9.4 x $10^{-13}$ ergs cm$^{-2}$ sec$^{-1}$ and an X-ray luminosity for the comet of 1.7 x $10^{14}$ ergs sec$^{-1}$, one of the fainter comets observed with *Chandra*. The light curve showed a dramatic decrease in flux of over 60% between observations on January 1$^{st}$ and 4$^{th}$. When comparing outer regions of the coma to inner regions, its spectra showed a decrease in ratios of CVI/CV, OVIII/OVII, as predicted by recent solar wind charge exchange emission models. There are remarkable differences between the X-ray emission from these two comets, further demonstrating the qualities of cometary X-ray observations, and solar wind charge exchange emission in more general as a means of remote diagnostic of the interaction of astrophysical plasmas.

# Acknowledgments

This work was supported by the Chandra X-ray observatory's cycle 9 GO program CXO-09100452 and archival program CXO-9100455. We thank Harvey Tannenbaum for Director's discretionary time to observe Comet 17P/Holmes (CXO-08108279), and we thank the Chandra X-ray Observatory's Scheduling and Mission Operations teams for executing these difficult and time critical moving target observations. S.J.W. was supported by NASA contract NAS8-03060. DJC thanks the California State University Northridge for support, and CML gratefully acknowledges support from Chandra GO program CXO-07108248. We also thank A. Fitzsimmons for useful discussions on 17P and Geronimo L. Villanueva for assistance in understanding the comet's orbital geometry. We are grateful for the cometary ephemerides of D. K. Yeomans published at the JPL/Horizons website.

# TABLES

**Table 1.** Observing parameters of comets 17P/2007 (Holmes) and 8P/2008 (Tuttle).

|  | 17P/2007 (Holmes) | 8P/2008 (Tuttle) |
|---|---|---|
| Observation Time (UT): | 2007 Oct 31.4 | 2008 Jan 01-04 |
| Exposure Time (ksec): | 30.0 | 46.8 |
| Heliocentric Distance (AU): | 2.46 | 1.10 |
| Geocentric Distance (AU): | 1.62 | 0.25 |
| Helio ecliptic Longitude (deg): | 54 | 87.9 |
| Helio ecliptic Latitude (deg): | +19 | +3.4 |
| Phase angle (deg): | 15 | 111.4 |

**Table 2.** Count rates for 17P using different background data. See text for details.

| (Background) Spectrum | Ra (deg) | Dec (deg) | l (deg) | b (deg) | Exposure Time (ksec) | Counts (300-1000 eV) | Counts (300-400 eV) | 17P Flux[1] $10^{-13}$ ph cm$^{-2}$ sec$^{-1}$ |
|---|---|---|---|---|---|---|---|---|
| Comet 17P/2007 (Holmes) | 56.7 | +50.5 | 149.11 | -3.26 | 30 | 4577 | 2121 | 10.0 |
| S3 Background |  |  |  |  | 30 | 930 | 333 | 4.3 |
| Obsid 4613 | 35.6 | +42.3 | 140.38 | -17.43 | 119 | 12762 | 3680 | 4.5 |
| Obsid 8473 | 49.36 | +41.41 | 150.30 | -13.60 | 30 | 11250 | 2033 | 1.0 |
| ACIS Blank Sky | - | - | … | … | 450 | 42815 | 16034 | 14.0 |

[1]Flux for 17P in the 300 to 1000 eV band, using this background in units of $10^{-13}$ ph cm$^{-2}$ sec$^{-1}$.

**Table 3.** Observed net counts for 8P's different spectral regions. See text for details.

| Region | Area (pix$^2$) | Counts (300-1000 eV) | Counts (300-500 eV) | Counts (500-700 eV) |
|---|---|---|---|---|
| Full Chip | 2.3x10$^6$ | 9330 | 3500 | 4040 |
| Nucleus | 3.1x10$^4$ | 530 | 220 | 260 |
| Panda0 | 6.3x10$^4$ | 870 | 360 | 420 |
| Panda1 | 6.3x10$^4$ | 630 | 240 | 304 |
| Panda2 | 6.3x10$^4$ | 520 | 190 | 240 |
| Left Panda | 6.3x10$^4$ | 840 | 340 | 390 |
| Background | 6.9x10$^5$ | 710 | 250 | 270 |

**Table 4.** Results of the spectral fit to the 17P/Holmes observations, with different backgrounds. See text for details.

| $E_{line}$ (eV) | Ion Line | Flux[1] 17P S3 BG | Flux[1] ID4613 BG | Flux[1,2] ID4613 | Flux[1] ID8473 |
|---|---|---|---|---|---|
| 299 | CV f+r+i | 930 ± 120 | 1100 ± 100 | 480 ± 130 | 160 ± 100 |
| 367.5 | CVI Ly-a | 10.5 ± 4.0 | 11.5 ± 5.0 | < 4.0 | < 5.0 |
| 419.8 | NVI f+r+i | < 7.0 | < 6.0 | … | … |
| 500.3 | NVII Ly-a | 1.9 ± 1.5 | 2.7 ± 0.9 | … | … |
| 561.1 | OVII f+r+i | < 1.3 | < 1.5 | … | … |
| 653.5 | OVIII Ly-a | <1.0 | < 0.9 | … | … |
| 907 | NeIX f+r+i | < 0.8 | < 0.8 | … | … |
| 1024 | NeX | < 0.9 | < 0.8 | … | … |
| $\chi^2$/dof | | 0.6 | 1.1 | 0.7 | 1.2 |
| Total Flux ($10^{-13}$ ergs cm$^{-2}$ s$^{-1}$) | | 4.3 | 4.5 | 1.6 | 0.5 |

[1] Line Fluxes in units of $10^{-5}$ ph cm$^{-2}$ sec$^{-1}$
[2] Background multiplied by a factor of 1.5

**Table 5.** Results of the spectral fit to the 8P/Tuttle observations for different extraction regions.

| $E_{line}$ (eV) | Line[1] | Full Chip | Panda0 | Panda1 | Panda2 | Left Panda | Nucleus |
|---|---|---|---|---|---|---|---|
| 299 | CV f+r+i | 4570 ± 700 | 442 ± 100 | 686 ± 101 | 482 ± 133 | 376 ± 120 | 197 ± 92 |
| 367.5 | CVI Ly-a | 209 ± 90 | 23 ± 16 | <16 | 41 ± 16 | 37 ± 15 | 25 ± 13 |
| 419.8 | NVI f+r+i | 128 ± 44 | 17 ± 8 | 29 ± 7 | < 10 | < 9 | < 6 |
| 500.3 | NVII Ly-a | 53 ± 16 | 7 ± 3 | 4 ± 3 | 8 ± 3 | 9 ± 3 | 5 ± 3 |
| 561.1 | OVII f+r+i | 165 ± 11 | 21 ± 2 | 26 ± 3 | 22 ± 2 | 18 ± 2 | 13 ± 3 |
| 653.5 | OVIII Ly-a | 59 ± 6 | 4 ± 1 | 5 ± 1 | 11 ± 1 | 4 ± 1 | 2 ± 1 |
| 907 | NeIX f+r+i | 11 ± 2 | < 0.2 | 1.1 ± 0.2 | 1.3 ± 0.3 | 0.3 ± 0.2 | < 0.1 |
| 1024 | NeX | 15 ± 2 | 0.5 ± 0.2 | 0.8 ± 0.2 | 1.8 ± 0.3 | 0.4 ± 0.2 | 0.3 ± 0.2 |
| $\chi^2$/dof | | 1.2 | 1.3 | 0.9 | 1.3 | 1.1 | 1.0 |
| Total Flux[2] ($10^{-13}$ ergs cm$^{-2}$ s$^{-1}$) | | 9.36 | 1.00 | 1.17 | 1.18 | 0.96 | 0.61 |

[1] Line Fluxes in units of $10^{-6}$ ph cm$^{-2}$ s$^{-1}$

[2] Flux for 8P in units of $10^{-13}$ ergs cm$^{-2}$ s$^{-1}$ in the 0.3 to 1.0 keV band.

**Table 6.** Results of the spectral fit to 8P/Tuttle observations ('full chip'), extracted at different time intervals

| $E_{line}$ (eV) | Line[1] | Interval | | | |
|---|---|---|---|---|---|
| | | A | B | C | Flare |
| 299 | CV f+r+i | 6480 ± 748 | 4483 ± 710 | 2847 ± 538 | 7572 ± 1257 |
| 367.5 | CVI Ly-a | 120 ± 110 | 100 ± 94 | <50 | <220 |
| 419.8 | NVI f+r+i | 187 ± 55 | 120 ± 52 | 89 ± 33 | 227 ± 84 |
| 500.3 | NVII Ly-a | 95 ± 19 | 20 ± 16 | 32 ± 18 | 120 ± 37 |
| 561.1 | OVII f+r+i | 208 ± 20 | 104 ± 15 | 41 ± 12 | 251 ± 30 |
| 653.5 | OVIII Ly-a | 61 ± 7 | 53 ± 7 | 35 ± 8 | 71 ± 12 |
| 907 | NeIX f+r+i | 10 ± 2 | 12 ± 2 | 9 ± 2 | 11 ± 3 |
| 1024 | NeX | 13 ± 2 | 14 ± 3 | 12 ± 3 | 10 ± 3 |
| $\chi^2$/dof | | 0.87 | 1.42 | 1.01 | 0.60 |
| Total Flux[2] ($10^{-13}$ ergs cm$^{-2}$ s$^{-1}$) | | 10.7 | 6.5 | 3.7 | 12.3 |

[1] Line Fluxes in units of $10^{-6}$ ph cm$^{-2}$ sec$^{-1}$

[2] Flux for 8P in units of $10^{-13}$ ergs cm$^{-2}$ sec$^{-1}$ in the 0.3 to 1.0 keV band.

**Table 7.** Solar Wind conditions and selected parameters for 17P and 8P and comparison to comets with different solar wind conditions from Bodewits et al. 2007.

| Comet | $Q_{gas}$[1] | | Solar Wind | | |
|---|---|---|---|---|---|
| | ($10^{28}$ mol s$^{-1}$) | Type | $V_p$ (km s$^{-1}$) | $\Delta t$ (days) | |
| 8P | 1.4 | Flare/Quiet | 360 | -0.25 | |
| 17P | 30-50 | Flare/CIR | 400,700 | 5.2 - 7.4 | |
| C/1999 S4 (LS4) | 3 | ICME | 592 | 1.09 | |
| C/1999 T1 (McH) | 6-20 | CIR/Flare | 353 | 6.63 | |
| 2P/2003 | 0.7 | Flare/PS | 583 | -1.09 | |
| 9P/2005 | 0.9 | Quiet | 402 | -0.38 | |

**Table 8.** Solar wind abundances for 8P and 17P relative to $O^{7+}$, obtained from the CXE-model fitting and comparison to Linear S4 (LS4) and McNaight-Hartley (McH). See text for details.

| Ion | 8P (full chip) | 8P (Int A) | 17P[1] | LS4 | McH |
|---|---|---|---|---|---|
| $O^{8+}$ | 0.28±0.05 | 0.23±0.03 | <0.02 | 0.32±0.03 | 0.42±0.04 |
| $C^{6+}$ | 1.6±0.10 | 0.71±0.08 | 0.21±0.08 | 1.4±0.4 | 0.95±0.4 |
| $C^{5+}$ | 69±5 | 77±5 | 38±12 | 12±4 | 15±5 |
| $N^{7+}$ | 0.25±0.03 | 0.35±0.04 | 0.024±0.018 | 0.07±0.06 | 0.19±0.06 |
| $N^{6+}$ | 1.03±0.08 | 1.20±0.12 | <0.16 | 0.63±0.21 | 0.47±0.20 |
| $Ne^{10+}$ | 0.04±0.03 | 0.03±0.02 | <0.01 | 0.02±0.01 | 0.004±0.005 |

[1]Normalized 17P to LS4 $O^{7+}$ line flux

**FIGURES**

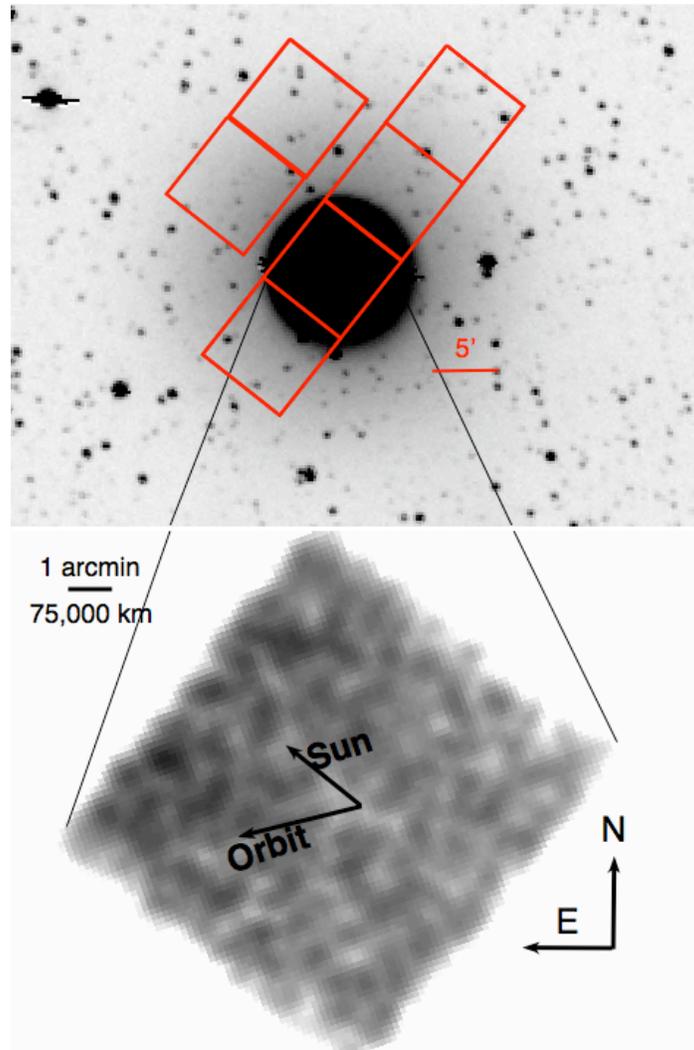

**Figure 1** - *Chandra* ACIS S3 pointing (bottom) compared to SuperWASP (top, V-band) observations of comet 17P/Holmes (Hsieh et al. 2007). The SuperWASP image was taken on Nov 01 at ~04:21 UT slightly after the *CHANDRA* image of Oct 31 10:07 UT. The SuperWASP pixel scale is ≈14"/pix.

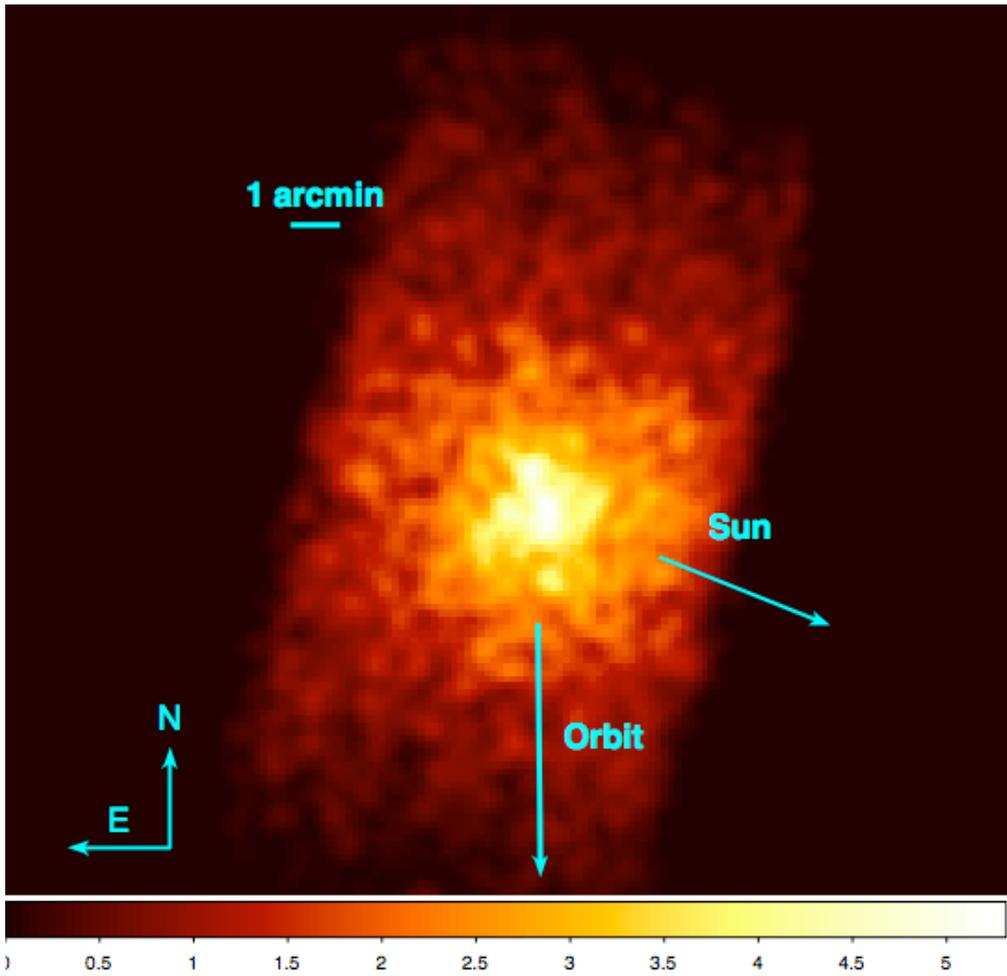

**Figure 2** - *Chandra* ACIS S3 300- 1000 eV image of 8P/Tuttle. North is up and East to the left and the position of the Sun and relative motion of the comet are indicated. The image is binned by 16 pixels and smoothed with a Gaussian filter of 3 pixels.

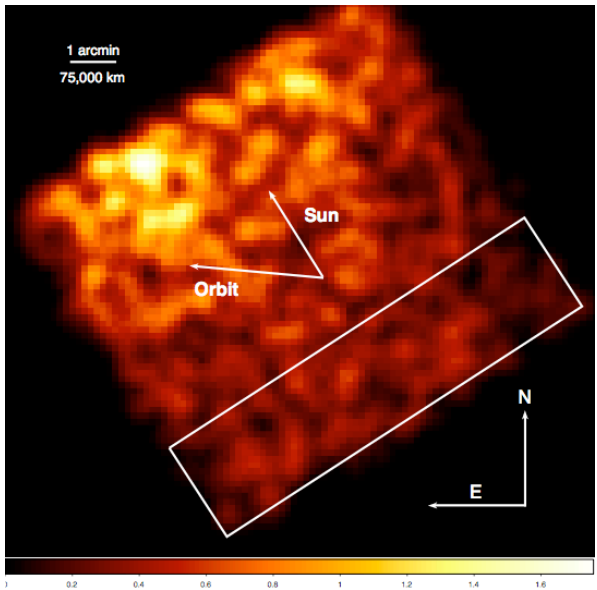

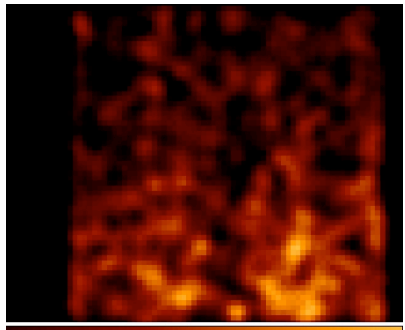 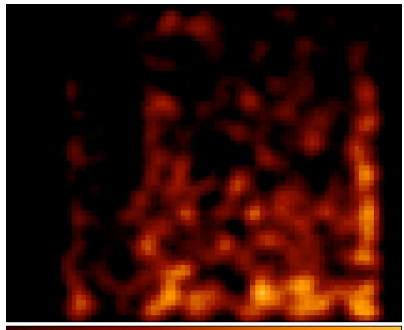 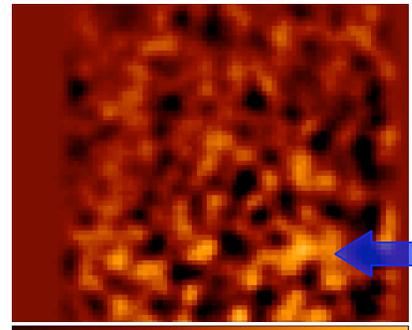

**Figure 3a** – *Chandra* image of comet 17P/2007 Holmes for the S3 CCD in the 300-400 eV range. The background region is indicated as the lower rectangle. **b-d.** *Chandra* images of comet 17P/2007 Holmes in detector coordinates in the 300-400 eV energy range, and comparison to the background image for ID8473. All images are the S3 CCD binned by 16 pixels and smoothed with a 3 pixel Gaussian filter and the CCD read-out is at the bottom. **b.** is the 17P image, **c.** is the ID 8473 background image and **d.** is the difference of the 17P and BG images. There is a small excess in the lower right side and this is marked with the blue arrow (see text).

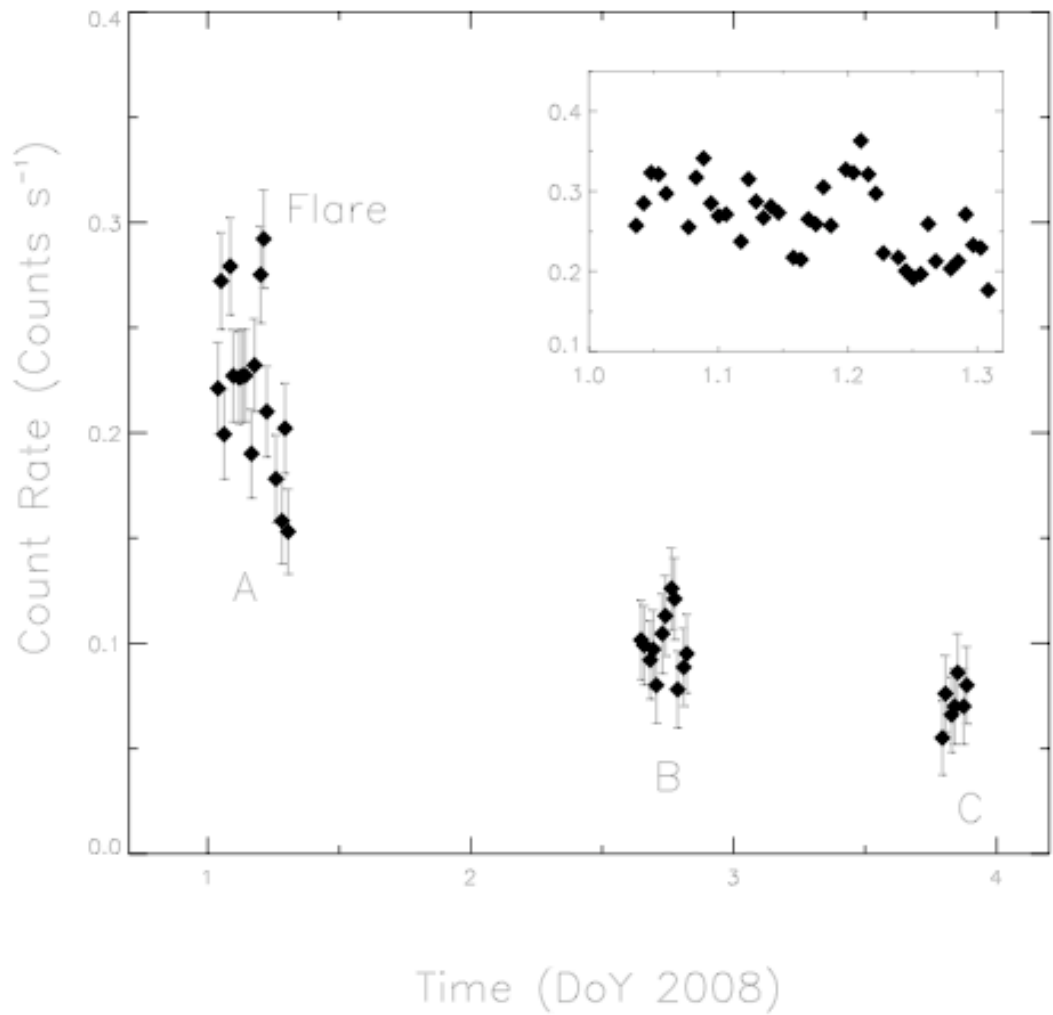

**Figure 4** – *Chandra* X-ray light curve of 8P shown in the 300 to 1000 eV range. The data are background subtracted and the bin size is 1000 seconds for the main plot. There was a modest flare observed on January 1st and this is expanded in the inset with a bin size of 500 seconds. The intervals used for spectral extraction are labeled. (see text).

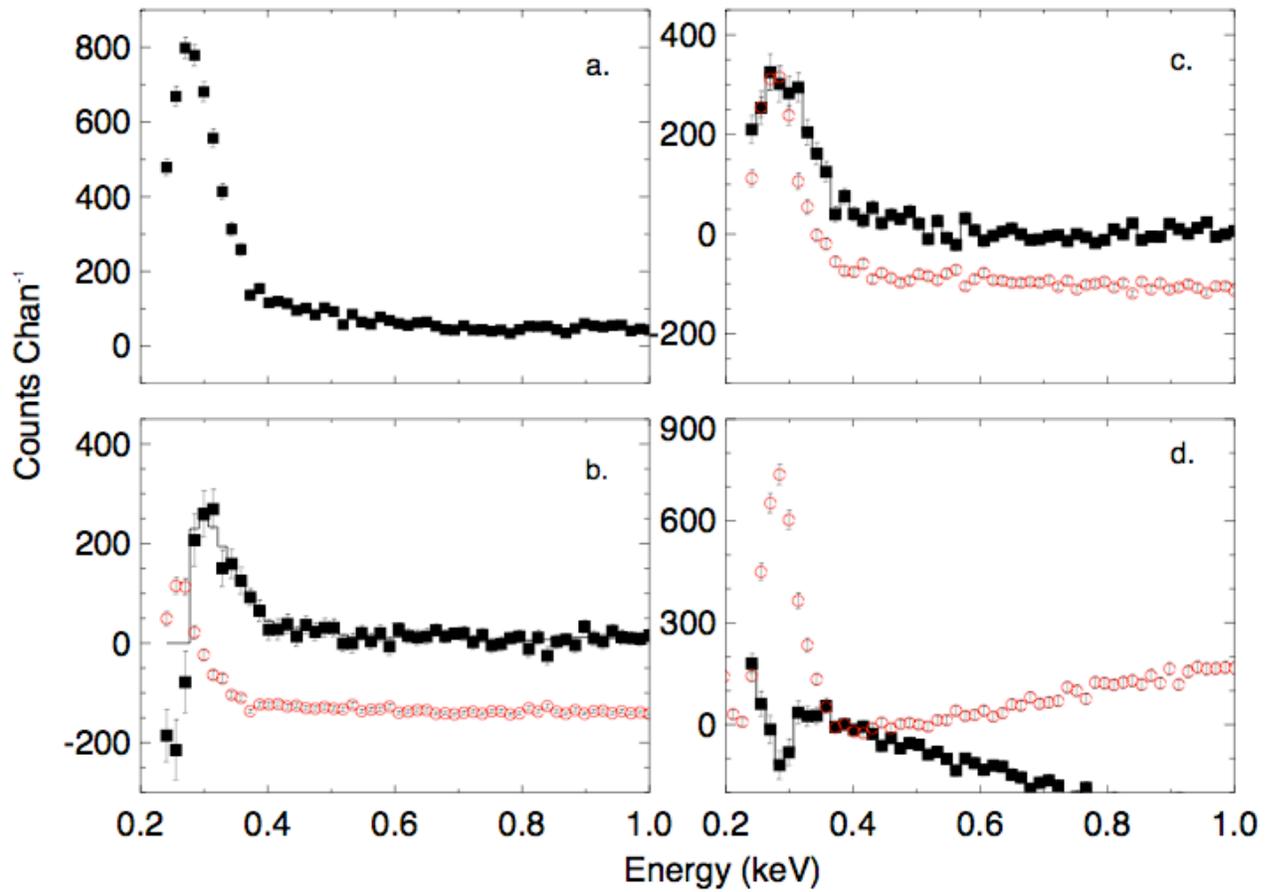

**Figure 5** – Comet 17P/Holmes ACIS S3 spectra. **a.** S3 full chip spectrum with no background subtracted (filled squares). **b** . S3 full chip spectrum subtracted with the S3 background spectrum extracted from lower section of the CCD. **c.** The 17P full chip spectrum subtracted with the ID4613 background. d. The S3 full chip spectrum subtracted from the ID8473 background in detector coordinates (see text). Over plotted in panels c-d are the background spectra offset by -150 counts (open red circles).

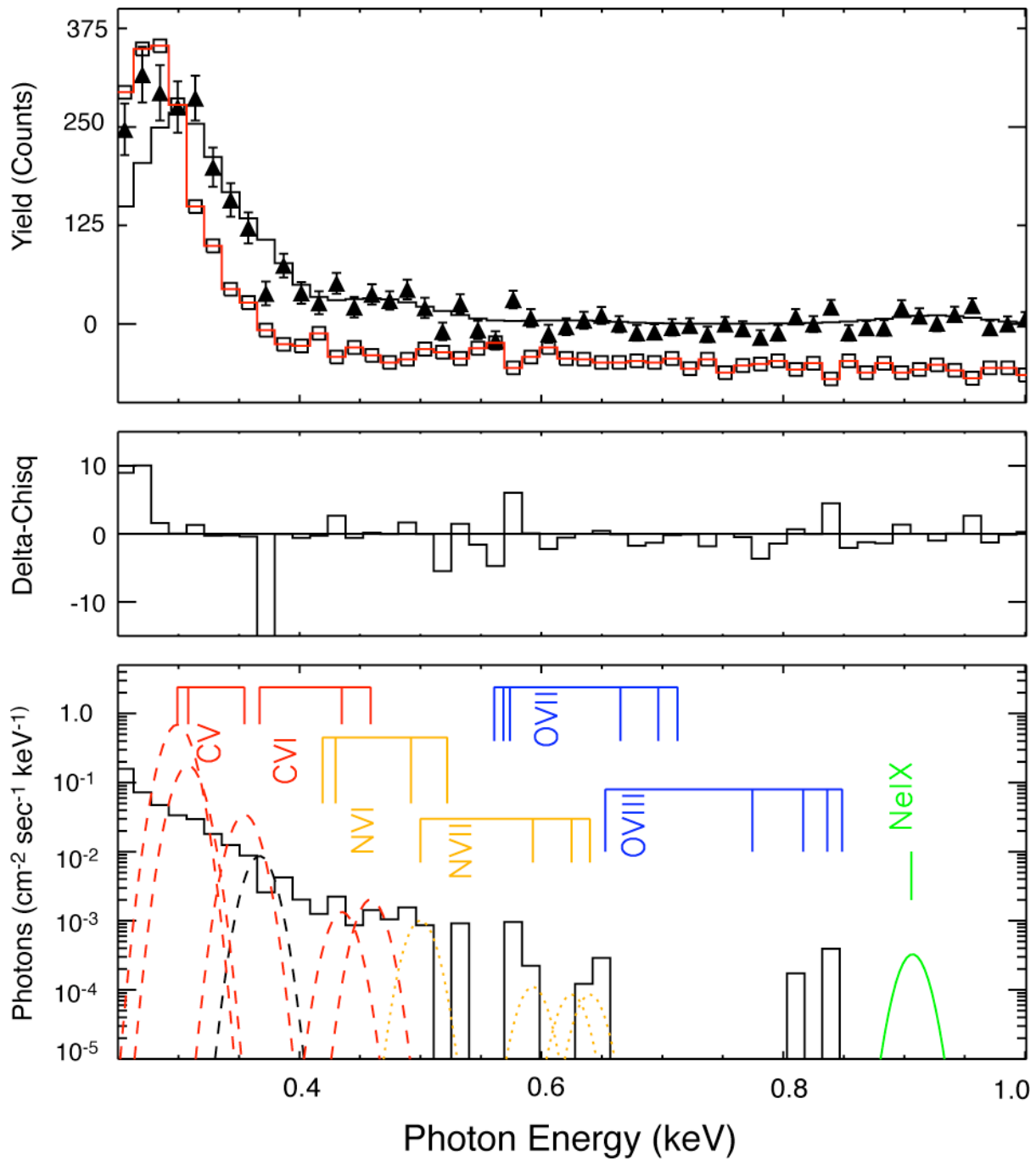

Figure 6 - Details of the SWCX fit for the spectrum of comet 17P/Holmes. **Top panel**: Background subtracted S3 full chip spectrum (filled triangles). The background (open squares, red line) used is the id4613 background (see text), scaled with a factor of 0.25, and offset downward by 100 counts for ease of presentation. **Middle panel:** Residuals of the SWCX fit. **Bottom panel:** SWCX 700 km/s model and observed spectrum indicating the different lines and their strengths. The unfolded model is scaled above the emission for the ease of presentation.

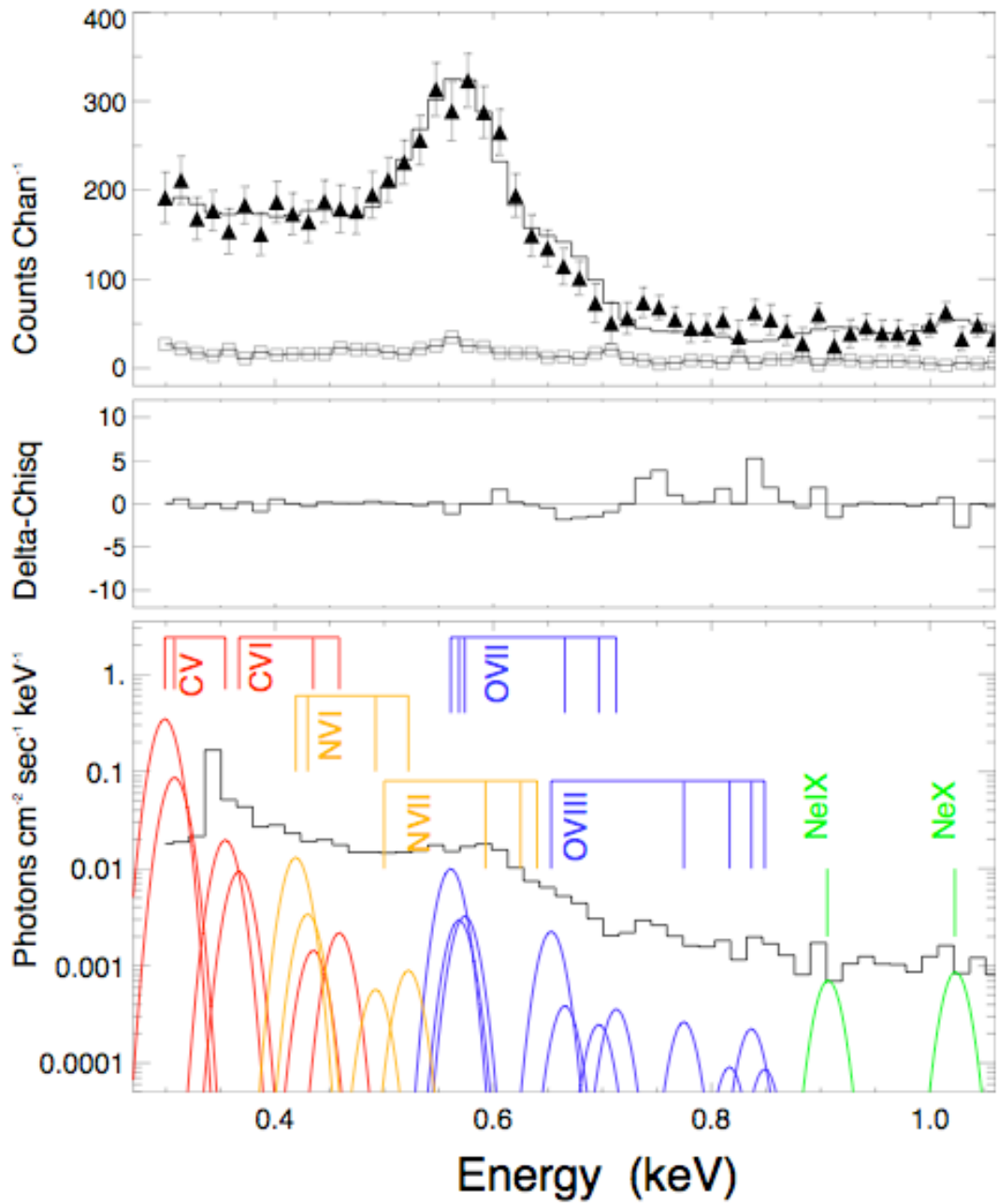

**Figure 7** - Details of the SWCX fit for the spectrum of comet 8P. **Top panel**: Background subtracted S3 full chip spectrum (filled triangles). The background (open squares) used. **Middle panel:** Residuals of the SWCX fit in units of $\chi^2$. **Bottom panel:** SWCX 400 km/s model and observed spectrum indicating the different lines and their strengths. The unfolded model is scaled above the emission for the ease of presentation.

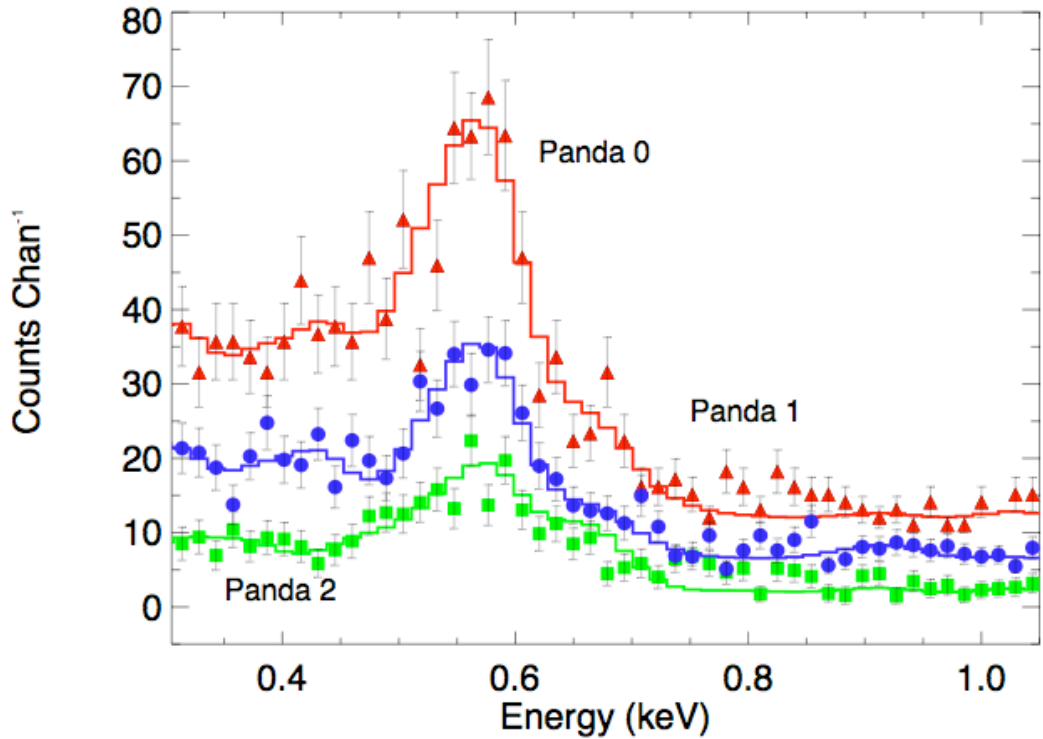

**Figure 8** - Comparison of spectra for the larger-scale structure of comet 8P for the entire observation. We compare three annular apertures (*Pandas*) centered on the nominal nucleus (see Section 2.2). Spectra are fit with the SWCX model with a solar wind velocity of 400 km s$^{-1}$. Pandas 1 (blue) and 2 (green) are scaled by their areas relative to Panda 0, and Panda 0 (red) and Panda 1 are offset above Panda 2, by 5 and 10 counts per channel, respectively for ease of presentation. All spectra show relatively strong emission from OVII and the inner most region (Panda 2, green) shows the strongest O VIII emission (see text).

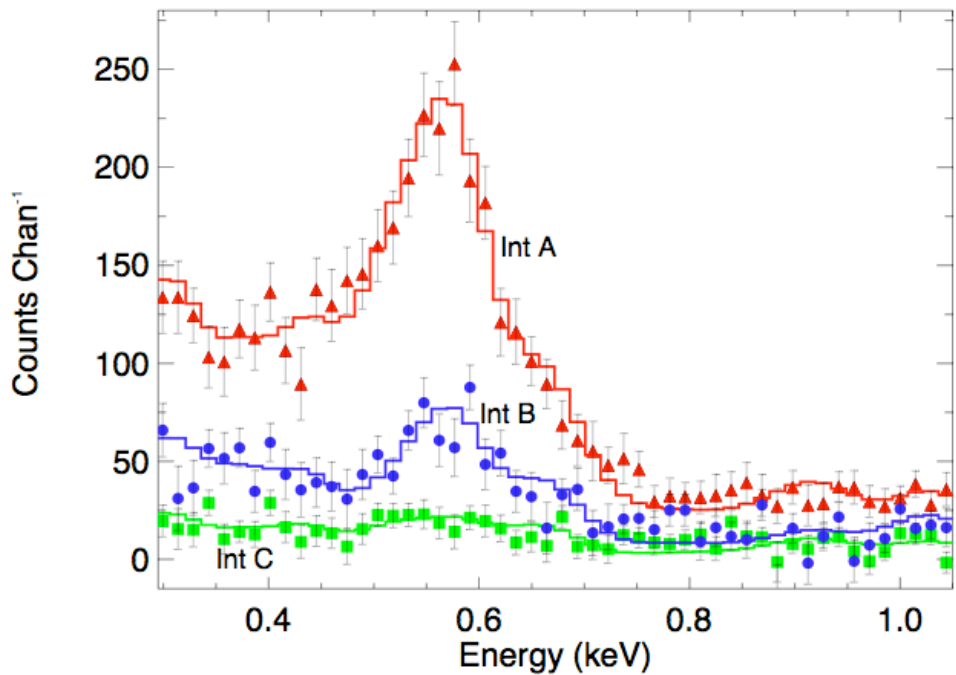

Figure 9 - Comparison of 8P full chip spectra as a function of time. Spectra are fit with the same SWCX model as figure 8, and shown are intervals A, B, and C as identified in Section 2.2. The spectrum of the small flare is not shown, but has a very similar shape as interval A, and has ~15% more flux (see text).

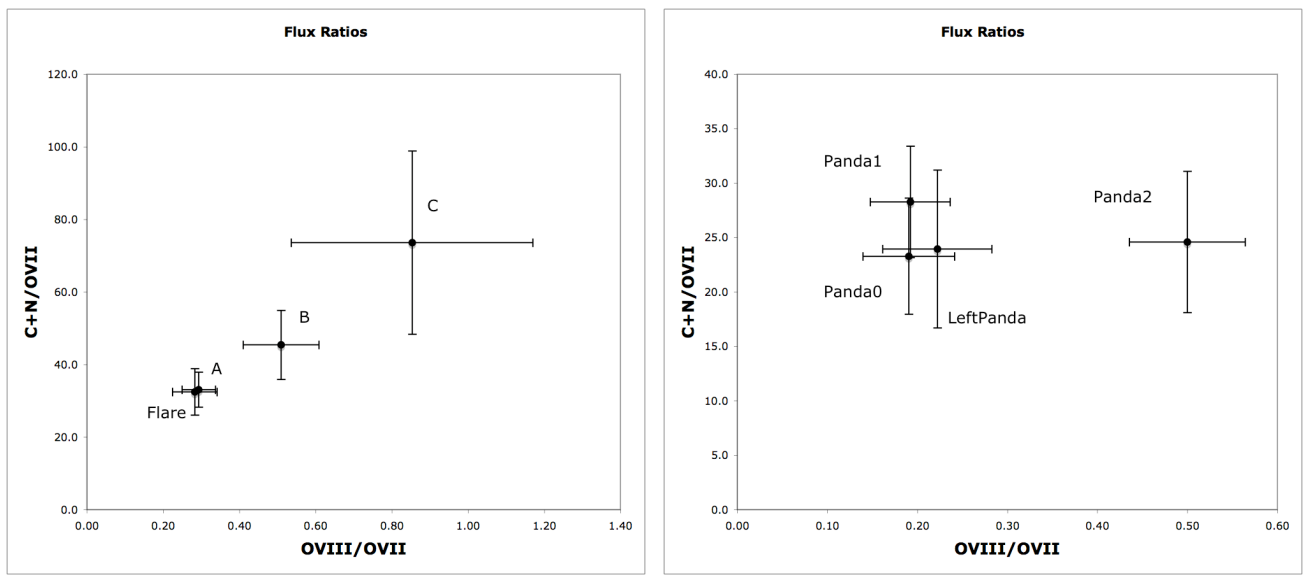

**Figure 10.** Emission line flux ratios from spectral fitting. Both figures show the ratio of the combined carbon plus nitrogen emission to that of the OVII emission plotted against the OVIII(653)/OVII(561) ratio. A. The left-hand figures shows these ratios for spectra extracted as a function of time. Each interval is labeled (A, B, C & Flare). Interval A and the flare spectra show the largest OVII emission and the lowest OVIII/OVII ratios. Interval C, the lowest flux interval shows similar amount of VII and OVIII and their ratio is nearly unity. B. The right-hand figure shows ratios from the spectral regions panda0,1,2 and the region behind the nucleus, noted as left panda. Here the region further from the nucleus shows the largest OVIII emission as expected from the simple SWCX model, but the combined carbon + nitrogen emission compared to the OVII emission stays nearly constant (see text).

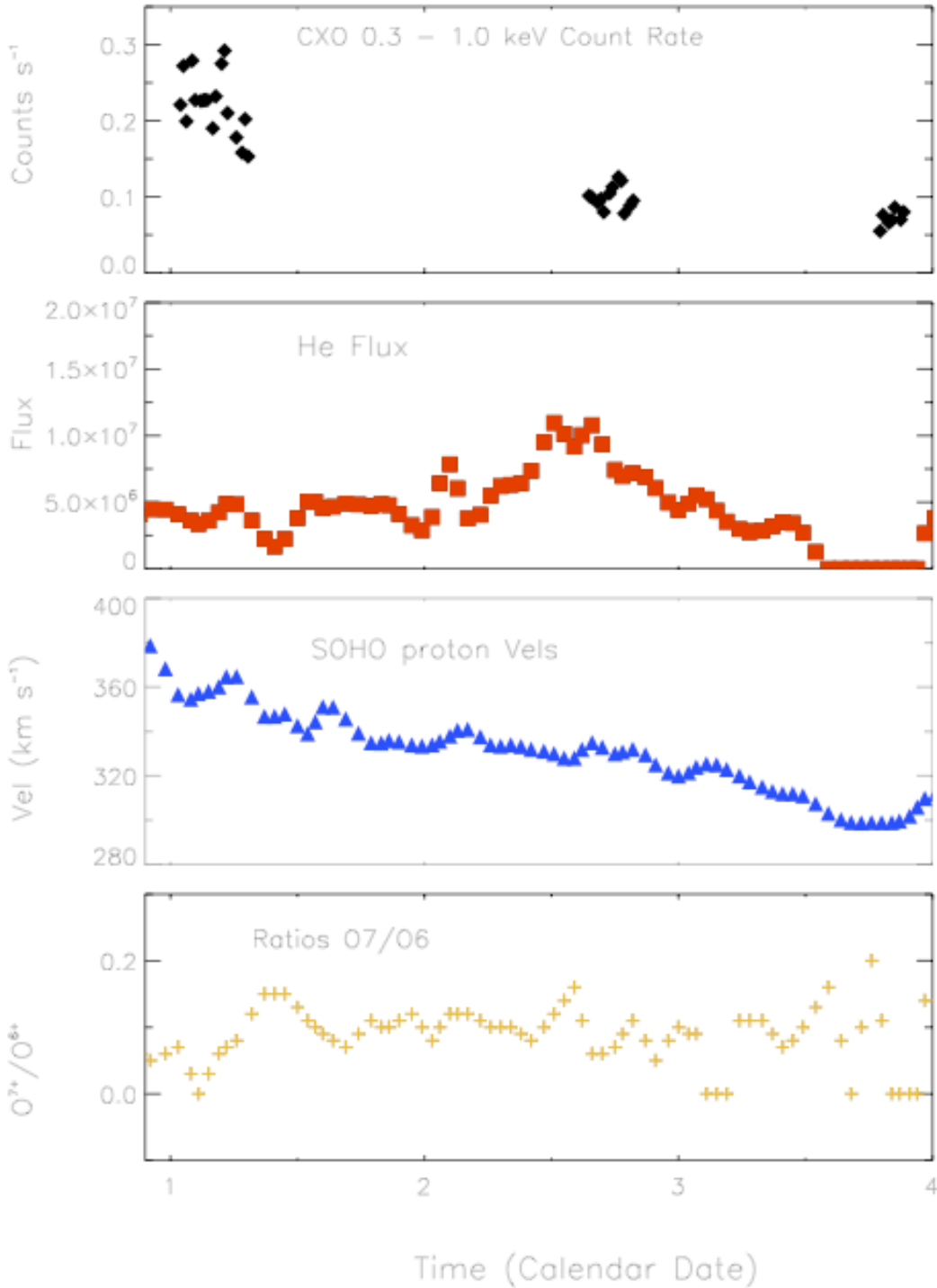

**Figure 11** - *Chandra* 8P light curve and measured solar wind parameters from ACE/SWICS and SOHO/PM. All data are mapped from the position of the spacecraft to that of the comet using a simple extrapolation for the solar wind (see Section 6.2). A) The top panel shows the *Chandra* X-ray light curve. B) Shows the $He^{2+}$ flux and C) the proton solar wind velocities from SOHO. The lowest panel, D) shows the $O^{7+}/O^{6+}$ (yellow points).